\documentclass[a4paper,12pt]{article}
\usepackage{epsf,rotate}

\newcommand{\be}{\begin{equation}}
\newcommand{\ee}{\end{equation}}
\newcommand{\bea}{\begin{eqnarray}}
\newcommand{\eea}{\end{eqnarray}}
\newcommand{\ben}{\begin{eqnarray*}}
\newcommand{\een}{\end{eqnarray*}}

\newcommand{\bra}[1]{\langle\, #1 |}
\newcommand{\ket}[1]{| #1\,\rangle}

\begin{document}

\begin{flushright}
hep-ph/9811475 \\
YUMS 98-021 \\
SNUTP 98-128 \\
KIAS-P98037
\end{flushright}

\vspace{1cm}

\centerline{\Large Determination of  HQET parameter $\lambda_1$}
\centerline{\Large from Inclusive Semileptonic $B$ Meson Decay Spectrum}

\bigskip

\begin{center}
K. K. Jeong$^{~a,}$\footnote{kkjeong@theory.yonsei.ac.kr} ~and~
C. S. Kim$^{~a,b,}$\footnote{kim@cskim.yonsei.ac.kr,
            ~~http://phya.yonsei.ac.kr/\~{}cskim/} \\[1cm]

$^a$ \it Department of Physics, Yonsei University, Seoul 120-749, Korea \\
$^b$ \it School of Physics, Korea Institute for Advanced Study, 
            Seoul 130-012, Korea
\end{center}

\vspace{1.5cm}

\begin{abstract}
We estimate the heavy quark effective theory parameter
$\lambda_1$ from inclusive semileptonic $B$-meson decay spectrum.
By using recent CLEO double lepton tagging data of $B \rightarrow X e \nu$, 
which shows the lepton
momentum as low as 0.6 GeV, we extracted $\lambda_1 \sim -0.58$ GeV$^2$.
We also derived $\overline\Lambda \sim 0.46$~GeV and
$|V_{cb}| = 0.041 \pm 0.002$.
\end{abstract}

\newpage

\section{Introduction}

As is well known, the heavy quark symmetry breaking parameters
$\lambda_1$ and $\lambda_2$ can affect the shape of
$B$ meson semileptonic decay spectrum substantially.
While it is easy to obtain the value of $\lambda_2$,
the hyperfine splitting term,
from mass difference between $B$ and $B^*$ mesons,
it is very difficult to determine the value of parameter $\lambda_1$,
which corresponds to the kinetic energy of heavy quark
inside a heavy meson.  So finding precise value of $\lambda_1$ is
very important in understanding of heavy meson decay.

The CLEO collaboration measured the lepton spectrum in the
inclusive $B\to X \ell\bar\nu$ decay both by one lepton tagging
\cite{barteltetal}, and by double lepton tagging \cite{cleo}.
In single lepton tagging data, leptons from secondary charm decay
($b\to c\to s\ell\nu$) dominate the low lepton energy region.
These secondary leptons have typically lower energy than the primary ones,
because they are from $c$ quark decay.
To obtain the $B\to X\ell\nu$ lepton spectrum in the low $E_\ell$ region
from the single lepton tagging data, these secondary leptons 
must be separated by fitting the spectrum with some assumptions and models.

In Ref.~\cite{gremmetal}, the parameter values of $\overline\Lambda$ and
$\lambda_1$ were estimated (with fixed value of 
$\lambda_2 = 0.12~\mbox{GeV}^2$) by using lepton energy distribution 
of $E_\ell> 1.5$~GeV from CLEO data~\cite{wang} of semileptonic decay 
$B \to X \ell \bar\nu$ with single lepton tagging. 
The advantage of using single lepton tagging data is small 
statistical error, though we cannot use low lepton energy part 
of the data ($E_\ell < 1.5$~GeV).

In Ref.~\cite{cleo}, CLEO collaboration separated
$B \to X \ell \bar\nu$ from cascade decays of $b\to c\to s\ell\nu$.
They selected events with tagging leptons of momentum greater than 1.4~GeV,
which are predominantly from semileptonic decay of one of the two
$B$ mesons in an $\Upsilon(4S)$ decay. When a tag was found, they searched
for an accompanying electron with minimum momentum 0.6~GeV.
The main sources of these electrons are, $(a)$ the secondary lepton from
the same $B$, $(b)$ the primary lepton from the other $B$ and
$(c)$ the secondary lepton from the other $B$.
Lepton from $(c)$ has the same charge as the tag lepton while
leptons from $(a)$ and $(b)$ have opposite charge to the tag lepton.
And leptons from $(a)$ and $(b)$ have different kinematic signatures so that
their contributions are easy to separate.
In the $\Upsilon(4S)$ decay, the $B$ and the $\overline B$ are produced
nearly at rest. Hence there is little correlation between the directions
of a tag lepton and an accompanying electron if they are from
different $B$ mesons. If they are from the same $B$, there is a tendency
for the tagged lepton and the electron to be back-to-back.
They analyzed the data with double lepton tagging and separated the primary
leptons from secondary leptons without model dependence.

In this paper by using the double lepton tagging data, we made a minimum 
$\chi^2$ analysis to determine value of the parameter $\lambda_1$.
There is one difficulty in $\chi^2$ fitting for the data, as is well known.
Since non-perturbative correction up to $1/m_b^2$ cannot predict
the correct shape of lepton distribution near the end point, 
we have to exclude 
the high energy data points of the distribution. 
Choosing $E_{QCD}$, the maximum lepton energy that one can trust 
the shape of $1/m_b$ expansion, is very important in this fitting.
Following Ref.~\cite{bigietal}, we choose $E_{QCD} = 2.0$~GeV.
The double tagged data has larger statistical error than the single
tagged one, but we can use low energy lepton data model-independently.
Therefore, this work can complement the work of Ref.~\cite{gremmetal}.

\section{Theoretical details}

Following the heavy quark effective theory (HQET)~\cite{eichten}, 
the mass of a pseudoscalar or a vector meson $M$ containing 
a heavy quark $Q$ can be expanded as 
\be \label{mass} 
m_M = m_Q + \overline\Lambda -
\frac{\lambda_1 + d_M \lambda_2}{2m_Q} + \cdots ,
\ee 
where $d_M=3,~-1$ for pseudoscalar and vector mesons, respectively, and
\bea 
\lambda_1 &=& \frac{1}{2m_M} \bra{M(v)} \, \bar h_v (iD)^2 h_v \ket{M(v)}, 
\\ 
\lambda_2 &=& \frac{1}{2 d_M m_M} \bra{M(v)} \, \bar h_v \frac{g}{2} 
\sigma_{\mu\nu} G^{\mu\nu} h_v \ket{M(v)}, 
\eea
where $h_v$ is the heavy quark field in the HQET with velocity $v$.
$\lambda_1$ parametrizes the mass shift due to the kinetic energy of 
heavy quark inside the meson, and $\lambda_2$ is related to the effect of 
chromomagnetic interaction between heavy quark and light 
degrees of freedom. 
In case of a $B$ meson, we can estimate the value of $\lambda_2$ quite 
accurately from the mass difference between 
$B$ and $B^*$ mesons. 
\bea 
m_{B^*} &=& m_b + \overline\Lambda - \frac{\lambda_1 - \lambda_2}{2m_b} , 
\\ 
m_B &=& m_b + \overline\Lambda - \frac{\lambda_1 + 3\lambda_2}{2m_b} , 
\label{Bmass}
\eea 
and approximately 
\be \label{lam2} 
\frac{1}{4}(m_{B^*}^2 - m_B^2) = \lambda_2 + \mathcal{O}(\frac{1}{m_b}) 
\approx 0.12~\mbox{GeV}^2 .
\ee 

Within HQET the lepton spectrum of the semileptonic decays of 
a $b$-flavored hadron $(H_b \to X_q\ell\nu )$ is calculated in 
the Ref.~\cite{bigietal,bloketal}, and the result is
\bea \label{hqet}
\frac{d\Gamma}{dx} &=& \Gamma_0 \theta (1-x-\epsilon^2) 2x^2 \Bigg[
(3-2x) - 3\epsilon^2 - \frac{3\epsilon^4}{(1-x)^2} + 
   \frac{(3-x)\epsilon^6}{(1-x)^3} \\
&& + G_b \left\{ \frac{6+5x}{3} - \frac{(6-4x)\epsilon^2}{(1-x)^2}
+ \frac{(3x-6)\epsilon^4}{(1-x)^3}
+ \frac{5(6-4x+x^2)\epsilon^6}{3(1-x)^4} \right\} \nonumber\\
&& + K_b \left\{ -\frac{5x}{3} + \frac{(2x^2 - 5x)\epsilon^4}{(1-x)^4}
+ \frac{2(x^3-5x^2+10x)\epsilon^6}{3(1-x)^5} \right\} \Bigg] , \nonumber
\eea
where
\begin{equation}
\Gamma_0 = \frac{G_F^2 m_b^5}{192\pi^3} |V_{qb}|^2 \quad , \quad
\epsilon = \frac{m_q}{m_b} \quad , \quad
x = 2E_\ell / m_b ,
\end{equation}
\be
K_b = -\lambda_1 / m_b^2 , \qquad G_b = 3\lambda_2 / m_b^2 ,
\ee
with $m_q$ denoting the mass of the quark $q=u,~c$ in the final state,
and $V_{qb}$ is CKM matrix element \cite{ckm}.
The terms in the second line and third line of Eq.~(\ref{hqet})
correspond to non-perturbative corrections (NP) to leading order 
Born approximation of the first line.

Perturbative corrections of the electron spectrum from 
$b$-decay were calculated in various 
references \cite{jezabeketal,perturbative}.
The analytic form of order $\alpha_s$ correction is 
given ~\cite{jezabeketal} as
\be
\left( \frac{d\Gamma}{dx} \right)_{\alpha_s}
= - \frac{2\alpha_s}{3\pi} \Gamma_0 
\int_0^{y_m} dy \frac{12}{(1-\xi y)^2 + \gamma^2} F_1(x , y, \epsilon^2) ,
\ee
with
\be
F_1(x, y, \epsilon^2) = H_1(x, y) + H_2(x, y, \epsilon^2) - H_2(x, y, z_m),
\ee
where
\bea
H_1(x,y) &=& 2(x-y)(x_M - x + y) H_B(x,y) \cr
&+& \frac{\bar Y_p}{2\bar p_3} \bigg\{ x(-4+5x) + y(4-6x-5x^2)
    + y^2(1+10x) - 5y^3 \cr
&+& \epsilon^2 [1-2x+5x^2 + y(5-16x) + 11y^2]
    + \epsilon^4(-2+6x-7y) + \epsilon^6 \bigg\} \cr
&+& \ln\epsilon [x(-1+2x) + y(1-4x) + 2y^2
    + \epsilon^2(1+x-y) - \epsilon^4],
\eea
and
\bea
H_2(x,y,z) &=& \frac{f_1 + zf_2}{8(1-y)[p_3(z)]^2}
    + \frac{Y_p(z) (f_3+zf_4)}{8[p_3(z)]^3}
    + \frac{Y_p(z) (f_5+zf_6)}{4p_3(z)} \cr
&+& \frac{1}{4} \ln(z) f_7 + \frac{\epsilon^2 f_8}{2(1-y)z}
    + [\mbox{Li}_2(w_+(z)) + \mbox{Li}_2(w_-(z))] f_9 \cr
&-& yz + 4yp_3(z) Y_p(z),
\eea
with
\bea
H_B(x,y) &=& 1-\ln(1-x) -\ln(1-y/x) - 2(\bar p_0/\bar p_3 \bar Y_p -1)
    \ln[(1-x)(1-y/x) -\epsilon^2] \cr
&+& \frac{\bar p_0}{\bar p_3} \bigg[ \mbox{Li}_2
    \left( 1-\frac{\bar p_- \bar w_-}{\bar p_+ \bar w_+} \right)
    -\mbox{Li}_2 \left( 1-\frac{\bar w_-}{\bar w_+} \right)
    -\mbox{Li}_2 \left( 1-\frac{\bar p_-}{\bar p_+} \right)
    +\mbox{Li}_2 \left( 1-\frac{1-x}{\bar p_+} \right) \cr
&+& \mbox{Li}_2 \left( 1-\frac{x-y}{x \bar p_+} \right)
    -\mbox{Li}_2 \left( 1-\frac{1-x}{\bar p_-} \right)
    -\mbox{Li}_2 \left( 1-\frac{x-y}{x \bar p_-} \right) \cr
&+& 2\bar Y_p (\bar Y_w + 2\ln\epsilon) + \ln\epsilon
\ln \frac{(\bar w_+ - x)(\bar w_+ - y/x)}{(x-\bar w_-)(y/x - \bar w_-)}
 \bigg].
\eea
And
\bea
f_1 &=& (1-y)^3[5x^2+y(5-2x)] - 4\epsilon^2(1-y)
    [x^2 + y(1-5x+2x^2) + y^2(2-x)] \cr
&&+ \epsilon^4 [-x^2 + y(-1+6x-3x^2) + y^2(-3+2x)], \\
f_2 &=& -(1-y)[5x^2 + y(5+18x+3x^2) + y^2(3-2x)] \cr
&&+ 4\epsilon^2 (1-y) [x^2+y(1-x)] + \epsilon^4 [x^2+y(1-2x)], \\
f_3 &=& (1-y)^2 [-5x^2 + y(-5-8x+x^2) + y^2] \cr
&&+ 2\epsilon^2 (1-y) [2x^2+y(2-6x+x^2) + y^2]
    + \epsilon^4 [x^2 + y(1-4x+x^2) + y^2], \\
f_4 &=& 5x^2 + y(5+28x+12x^2) + y^2(12+4x-x^2) - y^3 \cr
&&+ \epsilon^2 [-4x^2 + y(-4-4x+2x^2) + 2y^2] - \epsilon^4(x^2 + y), \\
f_5 &=& -5+10x +y(5+24x+8x^2) + y^2(5-18x) + 3y^2 \cr
&&+ \epsilon^2 [4-10x-4x^2 + y(-8+18x) -4y^2] + \epsilon^4(1-4x+y), \\
f_6 &=& 5+10x-4x^2 +y(14+10x) -3y^2 - 2\epsilon^2 (2+3x-2y) - \epsilon^4 ,\\
f_7 &=& -5+4x-4x^2+6yx-y^2 + \epsilon^2 [4(1+x)-10y] + \epsilon^4 , \\
f_8 &=& x(1-x) + y(-1+x+x^2) - 2y^2 x + y^3 + \epsilon^2 (1-x)(x-y) \\
f_9 &=& x+y(1-2x) + y^2 + \epsilon^2 (x-y).
\eea
All the parameters and the kinematic variables in the above expressions
are listed in Appendix.

After using the above all formulae, 
the electron distribution in semileptonic 
decay of $B$ meson can be written as
\be \label{dgamma}
\frac{d\Gamma_{theory}}{dE_\ell} 
= \left( \frac{d\Gamma}{dE_\ell} \right)_{Born}
+ \left( \frac{d\Gamma}{dE_\ell} \right)_{NP}
+ \left( \frac{d\Gamma}{dE_\ell} \right)_{\alpha_s} ,
\ee
where $\left( \frac{d\Gamma}{dE_\ell} \right)_{Born}$ is leading order
Born approximation,  $\left( \frac{d\Gamma}{dE_\ell} \right)_{NP}$
is non-perturbative correction using the HQET and
$\left( \frac{d\Gamma}{dE_\ell} \right)_{\alpha_s}$ 
the perturbative $\alpha_s$ correction.
We define the CKM-matrix independent decay rate
\be
\gamma_q = \frac{\Gamma_{theory}(B \to X_q \ell \nu)}{|V_{qb}|^2} ,
\ee
and then, semileptonic decay rate $\Gamma_{SL}$ can be written as 
\be
\Gamma_{SL} = \gamma_c |V_{cb}|^2 + \gamma_u |V_{ub}|^2 .
\ee
Since $|V_{ub}|^2 \ll |V_{cb}|^2$, we can neglect $b \to u$ decay.
Integrating over $E_\ell$ of Eq. (24), we obtain \cite{shifmanetal}
\be \label{totalgamma}
\Gamma_{SL} = \gamma_c |V_{cb}|^2
= \Gamma_0 \left[ z_0 \left\{ 1 -\frac{2\alpha_s(m_b)}{3\pi} 
g(\epsilon) \right\} + \frac{1}{2} z_0 (G_b - K_b) - 2 z_1 G_b \right] ,
\ee
where $z_0$ and $z_1$ are defined as
\bea
z_0 &=& 1 - 8\epsilon^2 + 8\epsilon^6 - \epsilon^8 - 
  24\epsilon^4 \ln\epsilon ~, \\
z_1 &=& (1-\epsilon^2)^4~,
\eea
and $g(\epsilon)$ is a complicated function of $\epsilon$, which can be 
approximated \cite{kimmartin} to 
\be
g(\epsilon) = \left(\pi^2 - \frac{31}{4} \right)(1-\epsilon)^2 + 
\frac{3}{2}~.
\ee
To obtain the mass ratio $\epsilon = m_c/m_b$, we use the relation
\bea \label{massdiff}
m_b - m_c &=& (m_B - m_D) - \frac{1}{2}(\lambda_1 + 3\lambda_2)
\left( \frac{1}{m_c} - \frac{1}{m_b} \right) \\
&\simeq& (m_B - m_D) \pm (\sim 1\%) \approx m_B - m_D = 3.41~\mbox{GeV} . 
\nonumber
\eea
We note that if we use instead the other relations, {\it e.g.}
\be
m_b - m_c = (m_{B^*} - m_{D^*}) - \frac{1}{2}(\lambda_1 - \lambda_2) 
\left( \frac{1}{m_c} - \frac{1}{m_b} \right)
\approx m_{B^*} - m_{D^*}=3.32~\mbox{GeV} ,
\ee 
or 
\bea 
m_b - m_c = (\overline{m}_B &-& \overline{m}_D) - \frac{\lambda_1}{2}  
\left( \frac{1}{m_c} - \frac{1}{m_b} \right)
\approx \overline{m}_B - \overline{m}_D=3.35~\mbox{GeV} , \\
{\rm where}~~~\overline{m}_B &=& \frac{1}{4}(m_B + 3 m_{B^*})~~~{\rm and}~~~
\overline{m}_D = \frac{1}{4}(m_D + 3 m_{D^*}) , \nonumber
\eea
the values of correction would become as large as $\sim +(2 \sim 4 \%)$ 
depending on  $\lambda_1$.

For $b \to c \ell \nu$, Figs. 1(a-b) illustrate the dependencies of various 
corrections on $m_b$ and $\lambda_1$. 
All figures in Fig. 1 are with $|V_{cb}| = 0.04$.
The value of $m_b$ determines mainly the overall size of decay width, 
while other parameters determine the shape of the distribution.
As can be seen from Figure 1(a), we find that the dependence 
of $\Gamma_{SL}$ 
on $m_b$ is rather weak on the contrary to
the naive estimation of $\Gamma_{SL} \propto m_b^5$, and is very sensitive 
to the quark mass difference $(m_b-m_c)$ \cite{shifmanetal}.
Note also that the shapes of Born approximation and 
perturbative correction are almost insensitive to the value of  $m_b$, 
while non-perturbative correction is quite sensitive to 
both $m_c/m_b$ and $\lambda_1$.

\section{Results and discussions}

To compare CLEO data with theoretical calculation,
we use minimum $\chi^2$ method with 
\begin{equation}
\chi^2 = \sum_{E_i < E_{QCD}} 
\frac{\left[ {\cal B}(E_i) - F^{\rm theory}(E_i;\epsilon,\lambda_1) 
 \right]^2}
{\sigma(E_i)^2} 
\end{equation}
where ${\cal B}(E_i)$ and $\sigma(E_i)$ are experimental 
data of differential 
branching ratio and error at lepton energy $E_i$, and
$F^{\rm theory}(E_i;\epsilon,\lambda_1)$ is theoretical 
prediction at $E_i$ as 
a function of parameters $\epsilon \equiv m_c/m_b$ and $\lambda_1$.
We normalized the decay distribution to have branching ratio of 10.49\% 
as in Ref.~\cite{cleo}, 
${\cal B}(B \to X\ell\nu) = (10.49 \pm 0.17 \pm 0.43)\%$, {\it i.e.},
\be
F^{\rm theory}(E_i;\epsilon \equiv m_c/m_b,\lambda_1) = 
\frac{0.1049}{\Gamma_{SL}}
\left[ \frac{d\Gamma_{theory}}{dE_\ell} \right]_{E_\ell = E_i} .
\ee
We note that, because of exact cancellation between
$\Gamma_{SL}$ and $\Gamma_{theory}$, $F^{\rm theory}$ is
independent of  $\Gamma_0$, and therefore independent of 
$|V_{cb}|$ and $m_b^5$.
$F^{\rm theory}$ is only indirectly dependent on $m_b$ through 
the definition
of $x$ in Eq. (8).
Following Ref. \cite{shifmanetal}, we use the $b$ quark mass,
$m_b = 4.8 \pm 0.1$ GeV,
which has been derived from QCD sum rule analysis of 
the $\Upsilon$ system \cite{voloshin}.
For $\alpha_s$, we use $\alpha_s = 0.22$, as in Ref.~\cite{gremmetal}.

We here comment on determination of $E_{QCD}$, which is crucial for this 
analysis.  
As we can see in the Figs. 1(b) and 2, non-perturbative corrections are 
significant only in large electron energy region 
({\it i.e.} $E_\ell > 1.5$~GeV).
Therefore, we have to include as many data points up to $E_{QCD}$, in which 
theory can give correct shape of lepton energy distribution. 
Otherwise, we cannot fully see the effect of non-perturbative correction 
which determines the value of $\lambda_1$. 
However, if we include the data points over $E_\ell > E_{QCD}$, 
the result will be meaningless
because the shape of the lepton energy spectrum is not reliable 
above $E_{QCD}$ region.
The numerical value of $E_{QCD}$  can be  estimated 
from the value of $m_b$ \cite{bigietal}:
For $b\to u$ decay with $m_u = 0$, 
$E_{QCD} \approx 0.9 \cdot m_b/2 \sim 2.15$~GeV.
For $b\to c$ decay, smaller smearing range is required near end point, and
\be 
 E_{QCD} \approx 0.9 \cdot (m_b^2 - m_c^2)/2 m_b \sim 2.0~{\rm GeV}.
\ee
Since we are dealing only with lepton energies less than 2.15~GeV,
we  neglect $b \to u$ decay and set 0.6 $< E_i <$ 2.0~GeV.

We tabulated the results in Table 1. 
Since we fixed $(m_b - m_c)$, changing the value 
of $m_b$ means changing mass 
ratio $\epsilon \equiv m_c/m_b$ together, 
and this mass ratio affects the results.
The values of 
$\overline\Lambda$ are determined from the mass relation Eq.~(\ref{mass}).
All values of $\lambda_1$ in Table 1 are much larger than the value 
in Ref.~\cite{gremmetal} 
which is $\lambda_1 = -0.19 \pm 0.10~\mbox{GeV}^2$, 
or the values in \cite{falketal,neubert} which are $\sim -0.1~\mbox{GeV}^2$,
but consistent with \cite{balletal,hwangetal,fazio} which
are in the range $-0.4 \sim -0.7~\mbox{GeV}^2$.

\begin{table}[b]
\caption{Results of the fitting with $m_b = 4.7 \sim 4.9$~GeV
and the fixed $m_b - m_c = 3.41$ GeV.}
\[
\begin{array}{cccc}
\hline
 m_b & \quad \epsilon^2 \equiv m_c^2/m_b^2 \quad& \lambda_1 & 
\overline\Lambda \\
\hline
4.7~\mbox{GeV} & 0.075 & -(0.45 \pm 0.19)~\mbox{GeV}^2 & 
0.57 \pm 0.018~\mbox{GeV} \\
\hline                                                            
4.8~\mbox{GeV} & 0.084 & -(0.58 \pm 0.23)~\mbox{GeV}^2 & 
0.46 \pm 0.022~\mbox{GeV} \\
\hline                                                            
4.9~\mbox{GeV} & 0.092 & -(0.70 \pm 0.27)~\mbox{GeV}^2 & 
0.34 \pm 0.026~\mbox{GeV} \\
\hline
\end{array}
\]
\end{table}

The values of $\lambda_1$ show significant dependencies on the input value 
of $m_b$, but still each value is consistent within $1 \sigma$ error range. 
As explained before, this large sensitivity comes from  mass ratio 
$m_c/m_b$.  Indeed, for $m_b = 4.8$~GeV, if we change the value of 
$(m_b - m_c)$ to 3.35 ~GeV,
{\it i.e.} $m_c^2/m_b^2 = 0.091$, 
then $\lambda_1 = -0.69 \pm 0.22~\mbox{GeV}^2$.
Changing the value of $m_b$ with fixed $m_c^2/m_b^2 = 0.084$,
we obtain $\lambda_1 = -0.55\pm 0.18~\mbox{GeV}^2$ for $m_b = 4.7$~GeV 
and $\lambda_1 = -0.52\pm 0.29~\mbox{GeV}^2$ for $m_b = 4.9$~GeV,
which are very similar to the case with $m_b = 4.8$~GeV, 
as shown in Table 1.

Once we know the parameter values $m_b$, $\lambda_1$, we can extract 
$|V_{cb}|$ 
from the relation
\be
|V_{cb}|^2 = \frac{{\cal B}(B \to X \ell \nu)}{\tau_B \gamma_c} ~. 
\ee
For $\tau_B$, we averaged $\tau_{B^\pm}$ and $\tau_{B^0}$ 
from Particle Data Book~\cite{PDBook}
\ben
\tau_{B^\pm} &=& (1.62\pm 0.06) \times 10^{-12} sec , \\
\tau_{B^0} &=& (1.56\pm 0.06) \times 10^{-12} sec .
\een
This gives the value
\be
|V_{cb}| = 0.041 \pm 0.002~,
\ee
where the error includes the errors from semileptonic branching ratio
of CLEO data, $B$ meson lifetime, uncertainties from $\lambda_1$ and 
$b$ quark mass.
This result is consistent with CLEO result with ISGW model which is
$|V_{cb}| = 0.040 \pm 0.001 \pm 0.002 $ \cite{cleo}, and with recent 
Particle Data Book result
$|V_{cb}| = 0.0395\pm 0.0017 $ \cite{PDBook}.

Figure 2 shows the best fit result of differential branching ratio 
compared with CLEO data as a function of charged lepton energy, 
with $m_b=4.8$ GeV and $\lambda_1= -0.58$ GeV$^2$.
It shows the relative size and shape of the various corrections for 
$m_b = 4.8$~GeV. Non-perturbative term is about $\sim -4.5\%$ 
and perturbative term is $\sim -12\%$ from leading approximation.
{}From these facts, it is clear that non-perturbative correction
determines the shape and not much effect on total decay rate, 
while perturbative term has little effect on the shape 
but its contribution on the total decay rate is quite large.
Fitting with data points between 1.0~GeV $< E_i <$ 2.0~GeV, 
we obtain $\lambda_1 = -0.57 \pm 0.19~\mbox{GeV}^2$ and 
$\overline\Lambda \simeq 0.46$ for $m_b = 4.8$~GeV, 
which are almost the same with the results from 0.6~GeV $< E_i <$ 2.0~GeV.
Finally we note the dependence on $\alpha_s$ is very weak. 
Changing $\alpha_s$ to 0.35 we get 
$\lambda_1 \sim -0.54~\mbox{GeV}^2$ and 
$\overline\Lambda \sim 0.46$~GeV for $m_b = 4.8$~GeV, 
which are almost same values in Table 1.

We finally note that recently
CLEO collaboration measured~\cite{cleo2} the first and the second 
moments of the hadronic mass-squared distribution in the inclusive 
decay $B \to X_c \ell\nu$ and also made a preliminary determination 
of the first and the second moments of the lepton energy distribution
from the spectrum in Ref.~\cite{cleo}.
Using those four moments, they obtained the values of $\bar\Lambda$ and 
$\lambda_1$, but there appeared to be inconsistencies in the results
which suggests either experimental error or problems in the HQET.
However, if we consider only the moments of the lepton energy distribution,
the preliminary CLEO analysis \cite{cleo2} gives
$\lambda_1 \sim -0.75 \pm 0.20~\mbox{GeV}^2$,
which is rather in a good agreement with our results.

\newpage

\centerline{\Large \bf Acknowledgments}
\medskip

We thank A. Falk for careful reading of the manuscript and his
valuable comments.
CSK wishes to thank the Korea Institute for Advanced Study for warm
hospitality.
The work KKJ was supported
in part by 1997-Non-Directed-Research-Fund, KRF,
in part by the CTP, Seoul National University,
in part by the BSRI Program, Ministry of Education, 
Project No. 98-015-D00061,
in part by the KOSEF-DFG large collaboration project,
Project No. 96-0702-01-01-2.
CSK wishes to acknowledge the financial
support of 1997-Sughak Program from Korean Research Foundation.


\newpage

\begin{appendix}
\section{Appendix -- Kinematic Variables}

In reference~\cite{jezabeketal}, the kinematic variables are defined as:
\begin{itemize}
\item $b, q, G, \ell, \nu$ : four-momenta of the $b$-quark,
 lighter quark, gluon, lepton, neutrino.

\item $P = q + G$, $W = \ell + \nu$ : 
four-momentum of the quark-gluon system and the virtual $W$

\item $\lambda_G$ stands for the scaled gluon mass
 ($\lambda_G \equiv m_G/m_b \ll \epsilon$)
\end{itemize}

The scaled masses and lepton energies
\be
\epsilon \equiv\frac{m_q}{m_b} =
    \left( \frac{q^2}{b^2} \right)^{\frac{1}{2}} , \quad
x \equiv \frac{2E_\ell}{m_b} ,\quad
y \equiv \frac{W^2}{b^2} ,\quad 
z \equiv \frac{P^2}{b^2} ,\quad
\xi = \frac{m_b^2}{M_W^2} ,\quad 
\gamma \equiv \frac{\Gamma_W}{M_W} 
\ee
vary in the region
\bea
0 \le & x & \le x_M \equiv 1 - \epsilon^2 \label{endpoint}\\
0 \le & y & \le y_m \equiv x(x_M - x)/(1-x) \\
(\epsilon + \lambda_G)^2 \le & z & \le z_m \equiv (1 - x)(1-y/x).
\eea
Frequently used kinematic variables which characterize the quark-gluon
system are
\bea
p_0(z) &\equiv& \textstyle \frac{1}{2} (1-y+z) , \cr
p_3(z) &\equiv& \textstyle \frac{1}{2} [1+y^2+z^2-2(y+z+yz)]^{1/2} , \cr
p_\pm(z) &\equiv& p_0(z) \pm p_3(z) , \cr
Y_p(z) &\equiv& \frac{1}{2} \ln\frac{p_+(z)}{p_-(z)} =
    \ln\frac{p_+(z)}{\sqrt{z}} ,
\eea
and similarly for the virtual $W$
\bea
w_0(z) &\equiv& \textstyle \frac{1}{2} (1+y-z) , \cr
w_3(z) &\equiv& \textstyle \frac{1}{2} [1+y^2+z^2-2(y+z+yz)]^{1/2} , \cr
w_\pm(z) &\equiv& w_0(z) \pm w_3(z) , \cr
Y_w(z) &\equiv& \frac{1}{2} \ln\frac{w_+(z)}{w_-(z)} =
    \ln\frac{w_+(z)}{\sqrt{z}} .
\eea
For $G=0$, which implies $z=\epsilon^2$, the abbreviations
\be
\begin{array}{ccc}
\bar p_0 \equiv p_0(\epsilon^2), & \quad \bar p_3 \equiv p_3(\epsilon^2), &
\quad {\it etc.,} \\
\bar w_0 \equiv w_0(\epsilon^2), & \quad \bar w_3 \equiv w_3(\epsilon^2), &
\quad {\it etc.}
\end{array}
\ee
will be useful. 
Polylogarithms are defined as real functions, and in particular
\begin{equation}
\mbox{Li}_2(x) = -\int_0^x \frac{dt}{t} \ln |1-t| .
\end{equation}
\end{appendix}

\newpage

\section*{Figure captions}
\begin{description}
\item[Fig. 1] 
Contributions of each terms in lepton spectra of $b\to c\ell\nu$ decay.
In all figures, $|V_{cb}| = 0.04$.
(a) Born approximation and perturbative $\alpha_s$ correction with 
$m_b = 4.7$~GeV (solid line), $m_b = 4.8$~GeV (dashed line), 
$m_b = 4.9$~GeV (dotted line) and $\alpha_s = 0.22$. 
(b) Non-perturbative correction with 
$\lambda_1 = 0.3~\mbox{GeV}^2$ (solid line),
$\lambda_1 = 0.4~\mbox{GeV}^2$ (dashed line),
$\lambda_1 = 0.5~\mbox{GeV}^2$ (dotted line).
$\lambda_2$ and $m_b$ are fixed with the values
$\lambda_2 = 0.12~\mbox{GeV}^2$ and $m_b = 4.8$~GeV.


\item[Fig. 2]
Best fit result for $m_b = 4.8~\mbox{GeV}$ with
Born approximation (long dashed line), non-perturbative correction (short
dashed line), perturbative correction (dotted line) and sum of
the all (solid line).
Dots with error bars represents CLEO data.
Parameter values are
$\lambda_1 = -0.58~\mbox{GeV}^2$, $\lambda_2 = 0.12~\mbox{GeV}^2$  and
$\alpha_s = 0.22$.

\end{description}

\begin{figure}[h]
\hspace{1.5cm}\rotate[r]{\epsfxsize=9cm \epsffile{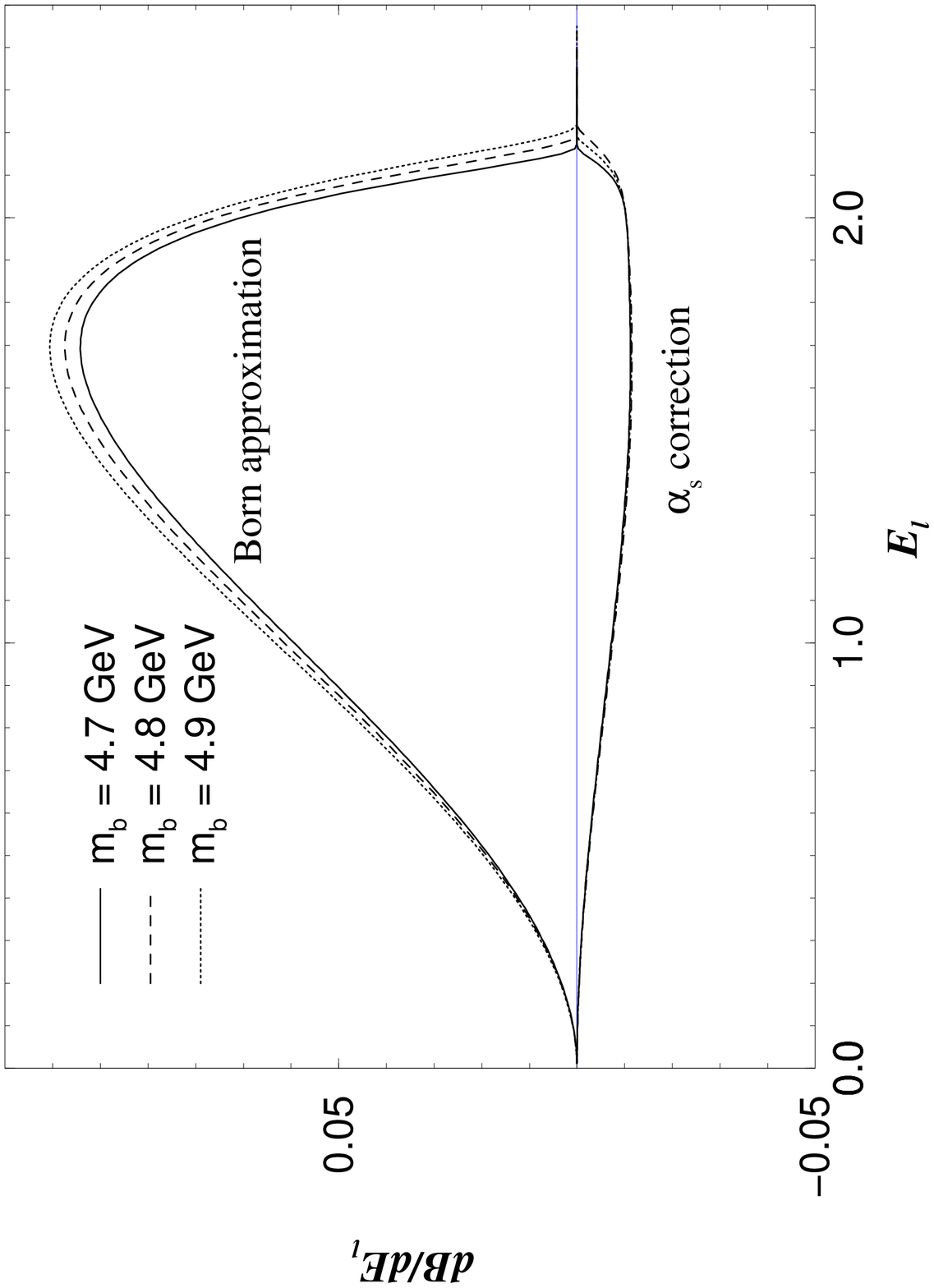}}
\caption{(a)}
\end{figure}
\setcounter{figure}{0}
\begin{figure}[t]
\hspace{1.5cm}\rotate[r]{\epsfxsize=9cm \epsffile{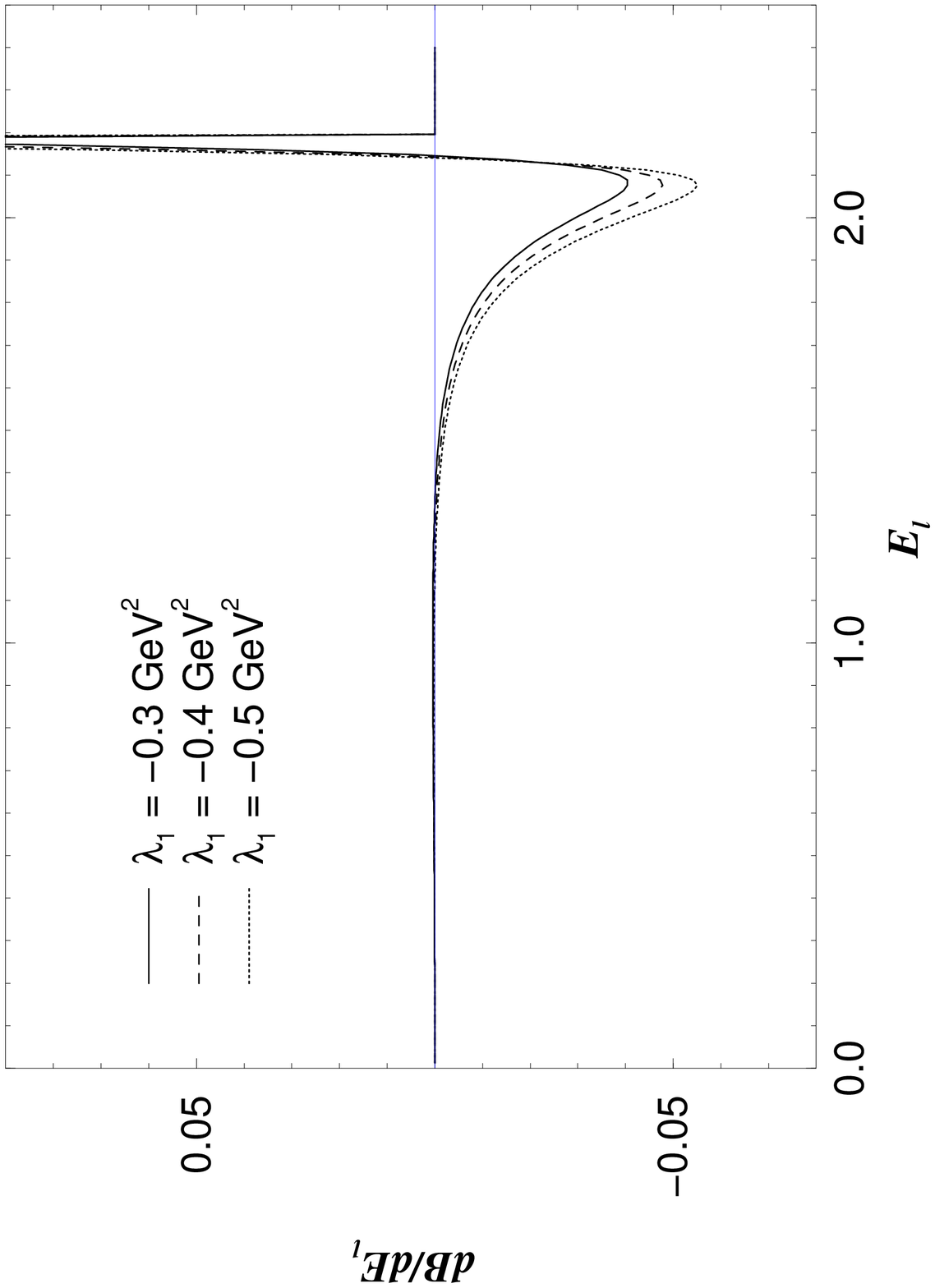}}
\caption{(b)}
\end{figure}
\begin{figure}[b]
\hspace{1.5cm}\rotate[r]{\epsfxsize=9cm \epsffile{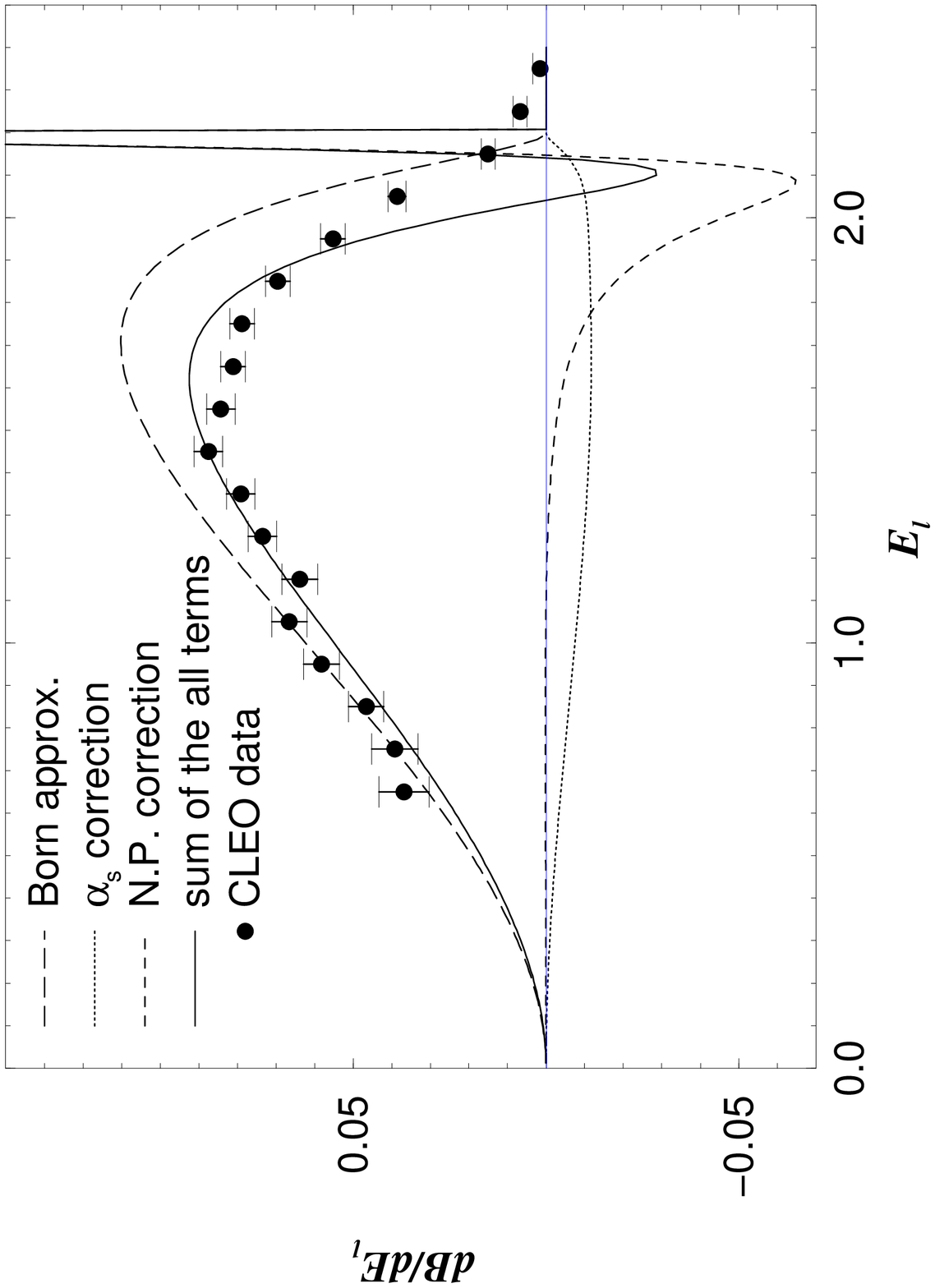}}
\caption{}
\end{figure}

\end{document}